\def\mearth{{\rm\,M_\oplus}}

\def\gsim{~\rlap{$>$}{\lower 1.0ex\hbox{$\sim$}}}
\def\lsim{~\rlap{$<$}{\lower 1.0ex\hbox{$\sim$}}}

\def\ctoo{\rm{[C] / [O]}}
\def\h2o{\rm{H_{2}O}}
\def\mh2{\rm{H_{2}}}
\def\c2h2{\rm{C_{2}H_{2}}}
\def\mh{\rm{H}}
\def\com{\rm{CO}}

\def\co2{\rm{CO_{2}}}
\def\ch4{\rm{CH_{4}}}
\documentclass[12pt, preprint]{aastex}
\usepackage{graphicx}
\usepackage{natbib,color,lscape}
\usepackage{subfigure}
\usepackage{threeparttable}
\usepackage{url}

\begin{document}
\title{A photochemical model for the carbon-rich planet WASP-12b}
\author{Ravi kumar Kopparapu\altaffilmark{1,2}, James F. Kasting\altaffilmark{1,2} and 
{Kevin J. Zahnle\altaffilmark{3}}}
\altaffiltext{1}{Department of Geosciences, Penn State University, 443 
Deike Building, University Park, PA 16802, USA}
\altaffiltext{2}{Virtual Planetary Laboratory}
\altaffiltext{3}{NASA Ames Research Center, MS 245-3, Moffett Field, CA 94035}

\begin{abstract}
The hot Jupiter WASP-12b is a heavily irradiated exoplanet in a short period 
orbit around a  G0-star with twice the metallicity of the Sun. A recent thermochemical 
equilibrium analysis based on Spitzer and ground-based infrared observations  suggests that
the presence of $\ch4$ in its atmosphere and the lack of $\h2o$ features can only be explained
if the  carbon-to-oxygen ratio in the planet's atmosphere is much greater than the solar ratio
($\ctoo = 0.54$). 
Here, we use a 1-D photochemical model to
study the effect of disequilibrium chemistry on the observed abundances of 
$\h2o, \com, \co2$ and $\ch4$ in the WASP-12b atmosphere. We consider two cases: one with 
solar $\ctoo$ and
another with $\ctoo = 1.08$. The solar case predicts that $\h2o$ and $\com$ are more abundant than
$\co2$ and $\ch4$, 
 as expected, whereas the high $\ctoo$ model
shows that $\com$, C$_{2}$H$_{2}$ and HCN are  more abundant.
This indicates that the extra carbon from the high $\ctoo$ model
is in hydrocarbon species.
$\h2o$ photolysis is the dominant disequilibrium 
mechanism that alters the chemistry at higher altitudes in the solar $\ctoo$ case,
 whereas photodissociation of C$_{2}$H$_{2}$ and HCN  is significant in the super-solar case.
Furthermore, our analysis indicates that $\c2h2$ is the major absorber in the atmosphere
of WASP-12b and the absorption features detected near $1.6$ and 8 micron
may be arising from C$_{2}$H$_{2}$ rather than $\ch4$. 
The Hubble Space Telescope's WFC3 can resolve this discrepancy, as $\c2h2$ has absorption between
$1.51 - 1.54$ microns, while $\ch4$ does not.


\end{abstract}

\maketitle

\section{Introduction}
\label{intro}
The discovery of the first  transiting planet, HD 209458b \citep{Charb2000,Henry2000},
opened up a new window to observe and study extrasolar planetary systems.
By combining transit data with radial velocity measurements, one can determine the
 mass and radius of a transiting planet 
\citep{Mazeh2000, Laughlin2005a, Laughlin2005b,HM2005,Agol2005}.
  Apart from these physical properties of the 
planet, it has also been shown that the transmission and emission spectra from 
ground- and space-based observations can be used to place constraints on the 
 atmospheric composition  \citep{Charb2002, VM2003, VM2004, Tinetti2007, Snellen2008,
Swain2009a, Swain2009b},
 brightness temperature \citep{Deming2005, Charb2005} and even day-night temperature contrast 
\citep{Knutson2007} of  transiting planets. To date, most of the planets discovered
are ``hot Jupiters''\citep{Collier2002}, but recently terrestrial mass planets are also
being discovered \citep{Leger2009,Charb2009}.

The observational determination of chemical species that exist in the outer atmospheric layers of
transiting planets provides us with an
opportunity to investigate the underlying chemistry. 
Typically, thermochemical equilibrium has been assumed in models of exoplanet atmospheres
\citep{Burrows1997,Fortney2005, Seager2005, Marley2007,Madhu2011}. This is a good assumption at 
the high temperatures and high pressures prevailing in the
lower layers of close-in gas giants.
Disequilibrium caused either by the UV flux of the host star (photochemistry) or  by
eddy and molecular diffusion (vertical transport) have been considered in some models
\citep{Liang2003, CS2006, Zahnle2009a,Zahnle2009b, line2010, Moses2011}. These studies
 showed that
disequilibrium mechanisms can be significant in determining the chemical composition of
hot Jupiters.

Here, we investigate the significance of photochemistry and transport in determining the abundances of 
major species, $\h2o,\com$ and $\ch4$, observed in the dayside thermal emission
spectrum \citep{Madhu2011, Croll2011} of  the transiting hot-Jupiter WASP-12b \citep{Hebb2009}.
At the time of its discovery, WASP-12b was the most highly irradiated exoplanet ( T$ > 2500$ K)
with the largest radius ($1.79 R_{J}$) and the shortest orbital period ($1.09$ days). 
Recently, \cite{Madhu2011} reported that Spitzer Space Telescope observations show
strong absorption features of $\ch4$ in the $3.6 ~\mu$m channel and $\com$ in the $4.5 ~\mu$m
channel, whereas weaker features were observed in the $5.8 ~\mu$m channel where $\h2o$ absorbs. 
This suggests that $\ch4$ and $\com$ are 
dominant and that $\h2o$ is less abundant in the atmosphere of WASP-12b. Assuming equilibrium
chemistry and solar $\ctoo = 0.54$, $\h2o$ and $\com$ should be the dominant species and
 $\ch4$ and $\co2$ should be the least abundant. Therefore, a solar $\ctoo$ ratio is ruled out.
\cite{Madhu2011} conclude that to explain the
observed abundance of $\ch4$ and $\com$, 
WASP-12b must have $\ctoo \ge 1$, implying that it is a carbon-rich planet. 

WASP-12b is one of the most highly irradiated known exoplanets, so photochemistry could play
 an important role in determining its atmospheric composition. 
Assuming$\ctoo = 1$, analysis of observations using equilibrium chemistry models suggests
mixing ratios (with respect to molecular hydrogen) 
 less than $ 10^{-7}$ for $\h2o$, greater than
$ 10^{-4}$ for $\com$, $\sim 10^{-5}$ for $\ch4$ and less than $ 10^{-9}$ for $\co2$.
Our goal in this study is to examine how the vertical distribution and  abundances 
of these species are affected by photochemistry. Specifically, we wish to determine whether
photochemical models make qualitatively different predictions from those of thermochemical
equilibrium models.
We consider two cases, one with solar $\ctoo = 0.54$ and another with $\ctoo = 1.08$, both of 
which have also been studied with equilibrium models.

\section{Model description}
\label{model}

We use a one-dimensional photochemical model initially developed to study primitive terrestrial 
atmospheres \citep{Kasting1982, Kasting1983, Zahnle1986, Kasting1990}. The model has been modified  to suit
the hot Jupiter temperature regime by including "backwards" chemical reactions that do not occur at the
low temperatures and pressures encountered in the Earth's atmosphere \citep{line2010,Moses2011}.
 This model solves a set of nonlinear,
coupled ordinary differential equations (ODEs) for the mixing ratios of all species  at all
heights using the reverse Euler method. The method is first order in time and uses second-order
centered finite differences in space. We include the following 
$31$ chemical species 
involved in $230$ reactions: O, O$_{2}$, $\h2o$, H, OH, $\co2$, $\com$, HCO, H$_{2}$CO,
$\ch4$, CH$_{3}$, CH$_{3}$O,CH$_{3}$OH, CH, CH$_{2}$,  H$_{2}$COH, C, C$_{2}$, C$_{2}$H,
C$_{2}$H$_{2}$, N, N$_{2}$, NH, NH$_{2}$, NH$_{3}$, CN, HCN,
H$_{2}$, He, O($^1$D) and $^{1}$CH$_{2}$. These species are divided into long-lived species
 (from O to H$_{2}$), 
short-lived species ( O($^1$D) and $^{1}$CH$_{2}$) and ``inert'' species (He). 
Both chemistry and vertical transport by eddy diffusion are considered for long-lived species,
whereas transport is neglected for short-lived species. Constant mixing ratios with altitude 
are assumed for
"inert" species. 
The
reaction list and rate constants were obtained from \cite{Zahnle2011} and 
are listed in Table \ref{table1}.  We have taken only the ``forward'' reactions and corresponding
rate coefficients, $k_{f}$, from \cite{Zahnle2011}: The reverse rate coefficients, $k_{r}$,
 at each temperature level (grid) were calculated assuming thermodynamic-equilibrium:
 $k_{r} = k_{f}/k_{eq}$, where $k_{eq}$ is the equilibrium constant for the reaction and is given by
$k_{eq} = e^{\Delta G^{\circ} /RT}$. Here $\Delta G^{\circ}$ is the change in the Gibbs free energy
 for the
reaction and is calculated from the Gibbs free energy of formation of reactants and products 
obtained from NIST-JANAF thermochemical tables when available\footnote{\url{http://webbook.nist.gov/chemistry/}} (and
from NASA thermobuild website\footnote{\url{http://www.grc.nasa.gov/WWW/CEAWeb/ceaThermoBuild.htm}}
when not available):
$\Delta G^{\circ} = \Delta G^{\circ}_{f} (reactants) - \Delta G^{\circ}_{f} (products)$.
However, it should be noted that 
one cannot simply calculate $k_{r}$ as discussed above. This is because the rate coefficients are 
given in units applicable to number densities whereas the thermodynamic quantities (enthalpy, 
entropy \& Gibbs energies) 
are calculated at a reference pressure (usually 1 bar). For reactions that have different number of
reactants and products, proper pressure terms must be added to obtain accurate rate coefficients.
We have appropriately included these terms in our reverse rates. Discussion of these
correction terms is given section $2.2$ of \cite{VM2011}.

 As lower boundary
conditions, we fix the mixing ratios of the species  at
 thermodynamic equilibrium values. 
Constant (zero) deposition velocities are assumed for the other species. The upper
boundary condition is set to zero flux for all the  long-lived species.
A more detailed description of the numerical scheme employed in this model 
is given in \cite{Pavlov2001}.

\begin{figure}[!hbp|t]
\includegraphics[width=0.92\textwidth]{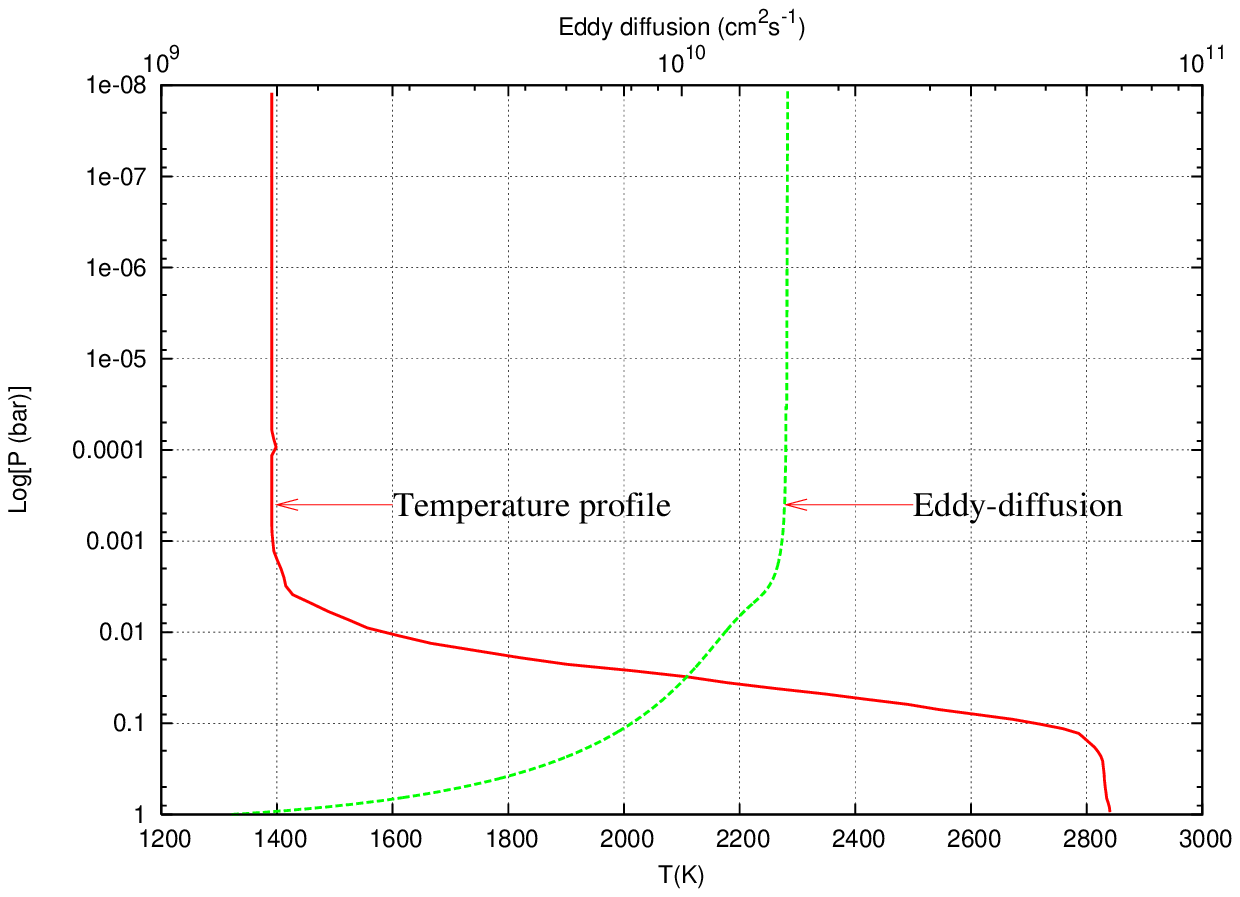}
\caption{Temperature profile (red solid curve) and eddy diffusion profile 
(green dashed line) used in our
photochemical model. The temperature profile is taken from one of the  models of 
\cite{Madhu2011} with no inversion, as a profile with inversion is ruled out by the data. 
This profile is then recalculated using hydrostatic equilibrium to be consistent with our
photochemical model.
Eddy diffusion profile is taken from the dayside profile of \cite{line2010}. }
\label{ptprofile}
\end{figure}

The vertical grid has $100$ altitude levels, ranging from $0$ km (lower boundary) 
to $12,800$ km (upper boundary) in $ 128$-km increments. The lower boundary pressure is set at
$1$ bar, and the upper boundary is fixed at $10^{-8}$ bar. Going to higher pressures is 
unnecessary, because the species profiles are already close to thermodynamic equilibrium well above
the 1-bar level. For the temperature profile, we
use one of the best-fit models from \cite{Madhu2011} which has no inversion
 (purple curve in their Fig. 1). The pressure profile was recalculated from this temperature profile
by assuming hydrostatic equilibrium and using the calculated mean molecular weights from the 
photochemical model. Vertical
transport is parameterized as eddy diffusion, as is common in one-dimensional photochemical models.
The dayside eddy diffusion profile from Fig. 1 of 
\cite{line2010}, which is originally obtained from the vertical winds from HD 189733b GCM of
\cite{showman2009}, is adopted. We have also performed sensitivity tests by varying eddy profiles, 
as discussed in \S\ref{discussion}. Both the
temperature and eddy profiles in our photochemical model are shown in Fig. \ref{ptprofile}.

For comparative purposes, we also calculate the thermodynamic equilibrium mixing ratios for all the
species in the photochemical model
at each altitude by solving simultaneously a system of chemical equilibrium equations.
These equations  require the total elemental abundances of carbon, oxygen,
hydrogen and nitrogen (as we consider only compounds from these elements) and Gibbs free energies as a
function of temperature. Solar elemental abundances from \cite{asplund2005} are assumed to be
our base values, but we report results for both solar $\ctoo$ and $2 \times$ solar $\ctoo$. To 
calculate Gibbs free energies, the enthalpy of formation at the reference temperature ($298$ K) and 
entropy are needed. 
  We then
use the expressions given in \cite{chase1998} (p. 16) to calculate Gibbs free energy of formation
for each species.

We initially tested our model by attempting to reproduce the results of  \cite{line2010},
for the hot Jupiter planet HD189733b. The dayside temperature and eddy diffusion profiles were 
taken from their Fig. 1. The lower boundary pressure was fixed at $10$ bar.
Both the thermochemical equilibrium and photochemical
model results are shown in  Fig. \ref{hd189733} of appendix \ref{supp} and
are in good agreement with the similar analyses of \cite{line2010} and \cite{Moses2011}. 
Moreover, our model
maintains equilibrium concentrations for all the major species in the deeper levels ($\sim 10$ bar),
 as it should at high temperatures and pressures.

\begin{figure}[!hbp|t]
\includegraphics[width=0.98\textwidth]{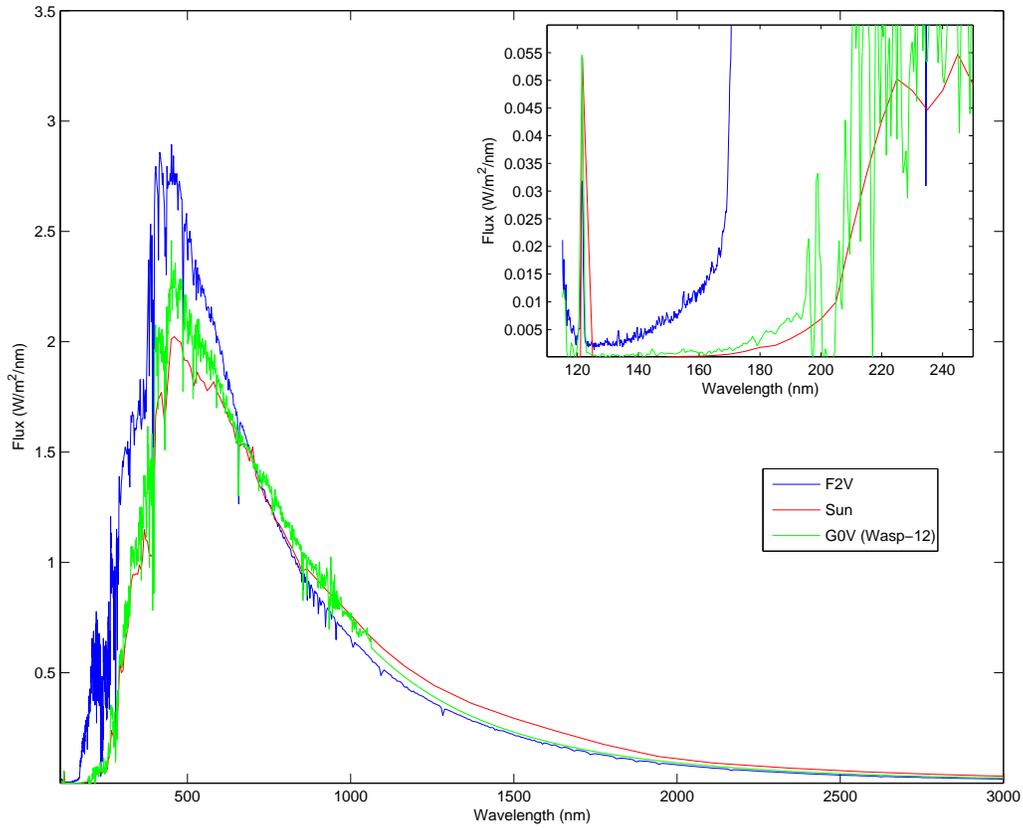}
\caption{Comparison of the normalized flux of a GOV stellar spectrum from \cite{Pickles1998}
 stellar
 flux library (green curve), which is assumed for the star WASP-12, with an F2V star
  (top blue curve) and the Sun (red bottom curve). Inset figure shows the UV region.}
\label{waspspectra}
\end{figure}

The star WASP-12 is a G0 star\footnote{\url{http://www.superwasp.org/wasp_planets.htm}} 
with an effective temperature of $6350$ K and twice the solar
metallicity \citep{Hebb2009}. To simulate its spectrum, we used a G0V star spectrum from
Pickles stellar spectral flux 
library \citep{Pickles1998}\footnote{\url{http://cdsarc.u-strasbg.fr/viz-bin/Cat?J/PASP/110/863}},
normalized to a solar flux of $1360$ W~m$^{-2}$ (the value at Earth's orbit today). 
We then multiplied the flux at each wavelength 
by a value consistent with inverse square law of the distance 
to get the correct flux for WASP-12b. The Pickles 
spectra are normalized to $1$ at 
$5556 \AA$. The fluxes from this dimensionless model spectrum are
 converted to W~m$^{-2}$nm$^{-1}$ by multiplying the following flux  expression 
from \cite{Gray1992}:
\begin{eqnarray}
\log F_{5556} &=& -0.40 V - 8.449
\end{eqnarray}
where $F_{5556}$ is the flux at $5556 \AA$ and $V$ is the visual magnitude of the star. For
WASP-12, $V = 11.69$ \citep{Hebb2009}.
In Fig. \ref{waspspectra}, we show the  G0V star spectrum along with F2V and the Sun.
A fixed stellar zenith angle of $50^{\circ}$ is assumed in all our models, the same value
that is used in our models of Earth's atmosphere (e.g. \cite{Pavlov2001}). This value is close
to the value of $48^{\circ}$ used by \cite{Moses2011} to reproduce secondary transit spectra in the 
atmosphere of HD 189733b.


\section{Results}
\label{results}

\begin{figure}[!hbp|t]
\includegraphics[width=0.92\textwidth]{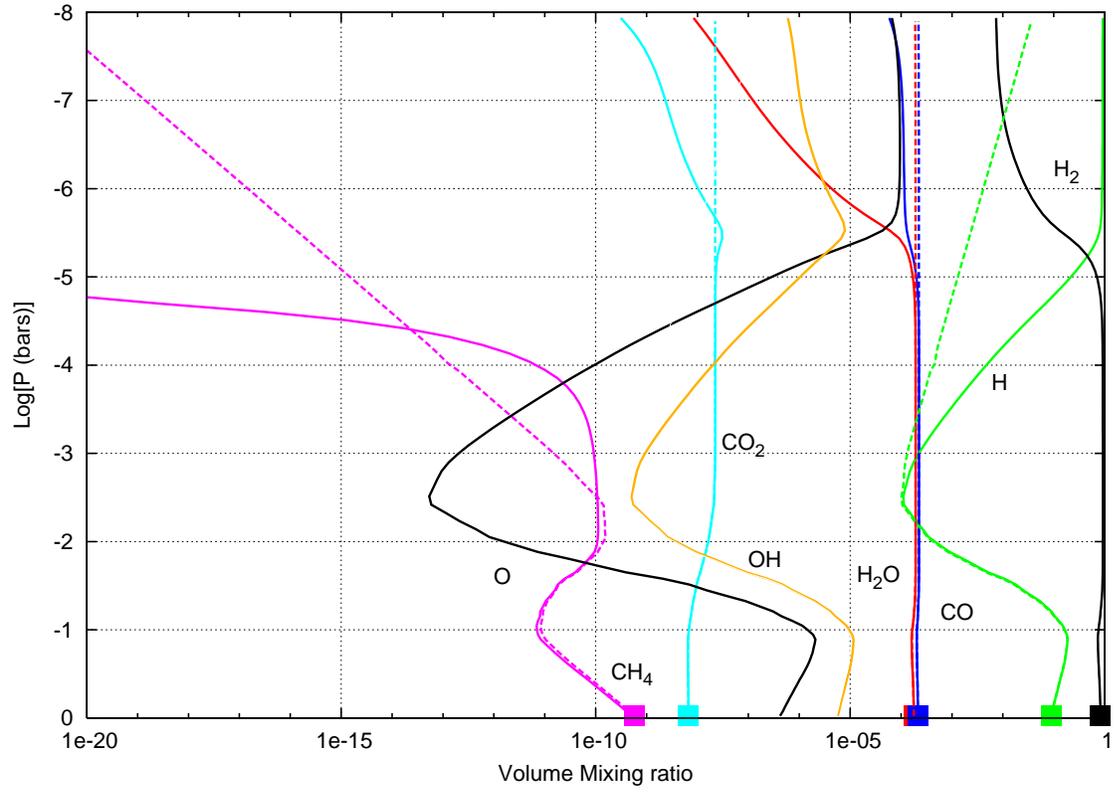}
\caption{Equilibrium (dashed) and photochemical (solid) mixing ratio profiles of
 major species, for $\ctoo = 0.54$ (the solar value). Filled squares represent equilibrium values 
at the lower boundary. The mixing ratios refer to volume mixing ratio. The Helium abundance is 
$0.07836$.}
\label{solar}
\end{figure}

We consider two different cases. In the first, we assume $\ctoo = 0.54$ (solar), and in 
the second we assume $\ctoo = 1.08$ (twice solar).
Fig. \ref{solar}  shows mixing ratio   profiles of  some of the major species 
 in our model, plotted against pressure 
for solar $\ctoo$ abundance. The lower boundary  in both the models is kept at
 $1$ bar pressure  ($T=2841 $ K) as
the observed spectral features mostly arise from pressures equal to or less than the 1-bar
level \citep{Fortney2005,Tinetti2007, Swain2009a, Madhu2011}.
Dashed lines represent the 
profile obtained from equilibrium chemistry, solid lines from our photochemical model
 and filled squares represent mixing-ratios of respective
species at the lower boundary.

In  the case of solar $\ctoo$ (Fig. \ref{solar}),  most of the
oxygen and carbon is in $\h2o$ and 
$\com$. The chemical loss time scale ($\tau_\mathrm{chem}$) for $\h2o, \com$ and
$\co2$ is smaller than the transport time scale ($\tau_\mathrm{trans}$); hence, as altitude increases
 the abundances stay at their equilibrium values until $\sim 10^{-5}$ bar \citep{PB1977}.
Below this pressure level (i.e., at higher altitudes) $\h2o$  gets photolyzed. 
$\h2o$ photolyzes at  lower altitudes than does $\com$  because the
dissociation energy for $\h2o$ ($5.17$ eV) is lower than that of $\com$  ($11.14$ eV)
(\cite{YD1999}, Table 2.4).
To break this strong C-O bond, photons of wavelength $\le 111.3$ nm are needed ($\h2o$
needs photons of wavelength $ \lsim 239.8$ to break its bond). $\com$ photolysis is not simulated
in our photochemical model and it can be a source of carbon and oxygen photochemistry at
high altitudes \citep{line2010, Moses2011}, but it should be relatively slow because of the
 small number of
photons at these short wavelengths. 
By contrast, the photon flux is quite high at the longer wavelengths that can photolyze $\h2o$
(green curve in Fig. \ref{waspspectra} inset).

In the case of $\co2$, the equilibrium abundance is maintained until 
$10^{-5}$ bar and is set by the following kinetic reactions that transfer  oxygen from 
$\com$ and $\h2o$ to $\co2$:
\begin{eqnarray}
\label{h2prod}
\mathrm{\h2o + H} \leftrightarrow \mathrm{H_{2} + OH} \\
\com + OH   \leftrightarrow \co2 + H
\label{co2prod}
\end{eqnarray}
At altitudes above $10^{-5}$ bar, $\h2o$ photolysis becomes the dominant source of OH production.
The OH  then combines with $\com$ through Eq. (\ref{co2prod}) to produce 
excess $\co2$ (local maximum of solid light-blue curve  $> 10^{-5}$ bar).
Above this level, $\co2$ becomes less abundant because it is itself photolyzed.

The shape of the equilibrium profile for  $\ch4$ can be understood from the following reaction
\begin{eqnarray}
\label{thermeqs1}
\com + 3 \mathrm{H_{2}} \leftrightarrow \ch4 + \h2o
\end{eqnarray}
and the corresponding equilibrium constant:
\begin{eqnarray}
\label{keq1}
K_\mathrm{eq} &=& \frac{p_{\ch4}~ p_{\h2o}}{p_{\com}~p_{\mh2}^{3}} 
\end{eqnarray}
where `$p_i$' represents the partial pressure of species $i$. The partial pressure is related 
to the total pressure and volume mixing ratio as $p_{i} = f_\mathrm{i} ~.~ P$.
As pressure increases (going downward) from $10^{-8}$ bar,
 the denominator term on the right-hand side of 
Eq.(\ref{keq1}) increases. Temperature is constant in this region (see Fig. \ref{ptprofile}), as is the
$\h2o / \com$ ratio. Thus, in order to maintain  equilibrium $\ch4$ must increase with depth.
Below $10^{-2}$ bar, the temperature starts to increase with depth.
 $\ch4$ is more stable at lower temperatures 
 and is also
more sensitive to temperature changes than other species.  Hence, it becomes less abundant
in the $10^{-2} - 10^{-1}$ bar regime. At pressures above $10^{-1}$ bar, the temperature again
  remains constant, so $\ch4$ must again increase with depth as it does in the upper atmosphere.

The $\ch4$ profile from the photochemical model
(solid magenta curve in Fig. \ref{solar}) follows the
 equilibrium profile at pressures up to $\approx  
10^{-2}$ bar. 
Above this level,
 $\tau_\mathrm{chem} \sim \tau_\mathrm{trans}$ (quench level) and 
 $\ch4$ remains well mixed near its equilibrium value of 
$10^{-10}$. Photolysis of $\ch4$ occurs above $10^{-3}$ bar (see Fig.\ref{solarphotorate}) mainly through the following reactions:
\begin{eqnarray}
\label{ch2}
\ch4 + h \nu &\rightarrow& \mathrm{CH_{2} + 2H}, \\
\label{ch3}
\ch4 + h \nu &\rightarrow& \mathrm{CH_{3} + H},\\
\label{ch21}
\ch4 + h \nu &\rightarrow&  \mathrm{^{1}CH_{2} + H_{2}}
\end{eqnarray}
Although H is produced through $\ch4$ photolysis, it is not enough to explain the increase in H abundance 
between $10^{-2} - 10^{-3}$ bar (light-blue solid curve in Fig.\ref{solar}). This increase in H is mainly due to the production
of OH through $\h2o$ photolysis at this level, which then combines with the most abundant
molecule in this atmosphere, $\mathrm{H_{2}}$, 
 through  the reverse of reaction (\ref{h2prod}).
\begin{eqnarray}
\label{h2ophot}
\h2o + h \nu &\rightarrow& \mathrm{H + OH}, \\
\label{h2oform}
\mathrm{H_{2} + OH} &\rightarrow& \h2o + H 
\end{eqnarray}

As the OH abundance increases, more $\mathrm{H_{2}}$ is consumed and it's mixing ratio decreases
above $10^{-5}$ bar (solid black curve in Fig. \ref{solar}). Eventually, $\h2o$ itself becomes
depleted by photolysis, so  the production of OH radical diminishes. At this point, $\mathrm{H_{2}}$
asymptotically reaches a mixing ratio of $10^{-2}$.
The increase in H also affects the atomic oxygen abundance (black solid curve in 
Fig. \ref{solar}) by the following reaction:
\begin{eqnarray}
\com + H \rightarrow \mathrm{O + CH}
\end{eqnarray}
Note that $\com$ photolysis, which is not included in our model, may dominate the above 
reaction in producing atomic oxygen.

\begin{figure}[!hbp|t]
\includegraphics[width=0.92\textwidth]{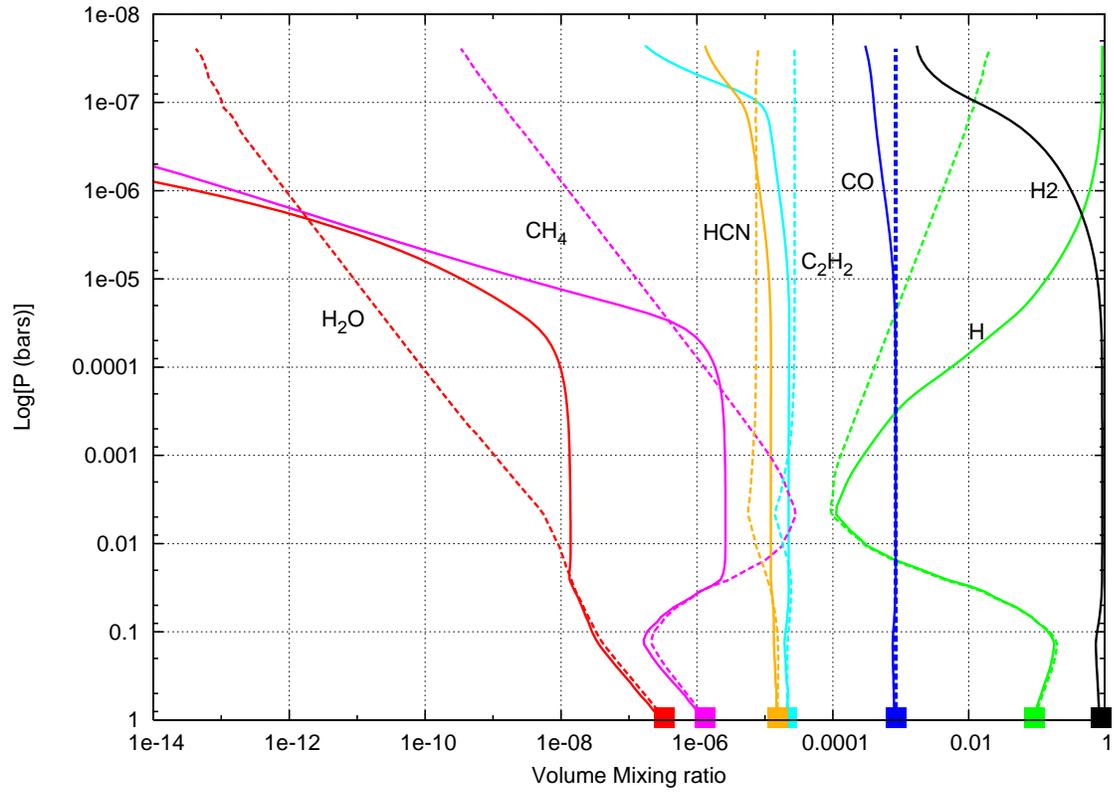}
\caption{Mixing ratio profiles from equilibrium (dashed) and photochemical (solid)
models for $\ctoo = 1.08$ (twice solar)  }
\label{solarCO1}
\end{figure}

Our analysis shows that, in the $\ctoo = 0.54$ case,
the abundances of major species (Fig. \ref{solar}) in WASP-12b's atmosphere are
mainly determined by thermochemical equilibrium, with departures at high altitudes due to 
disequilibrium chemistry driven by $\h2o$ photolysis. This is not surprising
considering that  $\h2o$ is  far more abundant than  $\ch4$.
The  photolysis rates of $\h2o$ and $\ch4$ as a function of pressure (altitude)
 for solar (blue) and
super-solar (red) $\ctoo$ are shown in
 Fig. \ref{solarphotorate}. In the solar case, at any given height,  $\h2o$ is more rapidly
 photodissociated than is $\ch4$, as it is  more abundant.
 In the super-solar $\ctoo$ case (Fig. \ref{solarCO1}), $\c2h2$ is more abundant than either $\h2o$ or
$\ch4$, and so it is photolyzed
 more rapidly at high
altitudes. 

\cite{Madhu2011} report that the spectrum obtained from the dayside multi-wavelength
 photometry of 
WASP-12b is best explained if one assumes  $\ctoo \ge 1$, using chemical
 equilibrium models. Under this 
assumption, the atmosphere is depleted in $\h2o$, enhanced in $\ch4$,
 and rich in $\com$. These 
equilibrium model profiles (dashed lines), along with our photochemical model results, 
for $\ctoo = 1.08 $ are shown in Fig. \ref{solarCO1}. In contrast to  
solar $\ctoo$ model, most of the oxygen is now in  
$\com$ (blue solid line), and $\ch4$ (magenta curve) is more abundant than $\h2o$ (red curve).
 The switchover from an atmosphere where $\h2o$ and $\com$ are the dominant species
 to one in which $\ch4$ and $\com$ become abundant happens precisely at $\ctoo = 1$. This transition is
 illustrated in
Fig. \ref{solarCO_transition}. 

In Fig. \ref{solarCO1} the abundance of $\h2o$ follows the equilibrium profile at pressures up to about
 $10^{-2}$ bar in the 
$\ctoo = 1.08$ model. At that point,
transport by eddy diffusion becomes faster than the chemical reaction timescale, so the equilibrium
value is maintained until $\sim 10^{-4}$ bar. Photolysis  begins above this level, and the
abundance of $\h2o$ decreases. The behavior of $\ch4$ is similar to the solar $\ctoo$ case, though
it is relatively more abundant in this high $\ctoo$ model. Note that for $\ch4$ and $\h2o$, the 
photochemical mixing ratios are not exactly equal to the equilibrium values below $0.1$ bar.
The reason is as follows: At high altitudes (above $10^{-6}$ bar), atomic hydrogen 
(green solid curve in Fig. \ref{solarCO1}) becomes a dominant species (more than H$_{2}$). Our 
photochemical model uses a minor constituent approximation for the diffusion coefficient 
in a binary mixture (Eq.(15.29), \cite{BK1973}), which  clearly is not applicable to H at this 
level. Due to this
approximation, the mixing ratio of H exceeds unity above $10^{-6}$ bar, which is 
unphysical. Therefore, we renormalize the mixing ratios of each species in our photochemical
model so that they sum to unity, and hence the equilibrium and photochemical profiles deviate slightly
in the lower atmosphere. Note that this should
not affect our conclusions in any way regarding which species are dominant (discussed in the next
paragraph) in Wasp-12b's high $\ctoo$ model.

As can be seen from Fig. \ref{solarCO1}, and also pointed by \cite{Moses2011b},
 the dominant hydrogen species 
(apart from H and H$_{2}$) in this model are HCN and $\c2h2$. Therefore, the
 photolysis of these two species drive the 
disequilibrium chemistry in the upper atmosphere. For example, in the solar model, the catalytic
H$_{2}$ destruction mechanism initiated by $\h2o$ photolysis (Eqs. (\ref{h2ophot}) \&
(\ref{h2oform})) was used to explain the
increase in H abundance shown in Fig. \ref{solar} (green solid curve).
 A similar increase of H at high altitudes can be noticed in the high $\ctoo$ case.
Reactions (\ref{h2ophot}) and (\ref{h2oform}) require OH production through $\h2o$ photolysis,
 which is negligible in the high $\ctoo$ model. Instead, the following reactions are important:

\begin{eqnarray}
\label{c2h2phot}
\c2h2 + h \nu &\rightarrow& \mathrm{C_{2}H + H}, \\
\label{c2h2prod}
\mathrm{C_{2}H + H_{2}} &\rightarrow& \c2h2 + H, \\
\label{hcnphot}
\mathrm{HCN + h \nu} &\rightarrow& \mathrm{CN + H}, \\
\label{hcnprod}
\mathrm{CN + H_{2}} &\rightarrow& \mathrm{HCN + H}
\end{eqnarray}
The results of these reactions can be seen in Fig. \ref{solarCO1}: At altitudes above $\sim 10^{-2}$ bar,
the photolysis of $\c2h2$ and HCN produce C$_{2}$H, CN and H through the above reactions. 
An increase in H can be seen as a result. The abundances of CN \& C$_{2}$H are not large 
enough below $10^{-5}$ bar to have a significant effect on the mixing ratio of H$_{2}$.
Above this level, reactions (\ref{c2h2prod}) \& (\ref{hcnprod}) result in the decrease of
H$_{2}$ mixing ratio (solid black line in Fig. \ref{solarCO1}) and corresponding increase of
H. Further up, $\c2h2$ and HCN become scarce and the production of C$_{2}$H and CN diminishes,
 which in turn reduces the rate of production of H. Therefore, H assumes a nearly constant mixing ratio 
thereafter.

\begin{figure}[!hbp|t]
\includegraphics[width=0.92\textwidth]{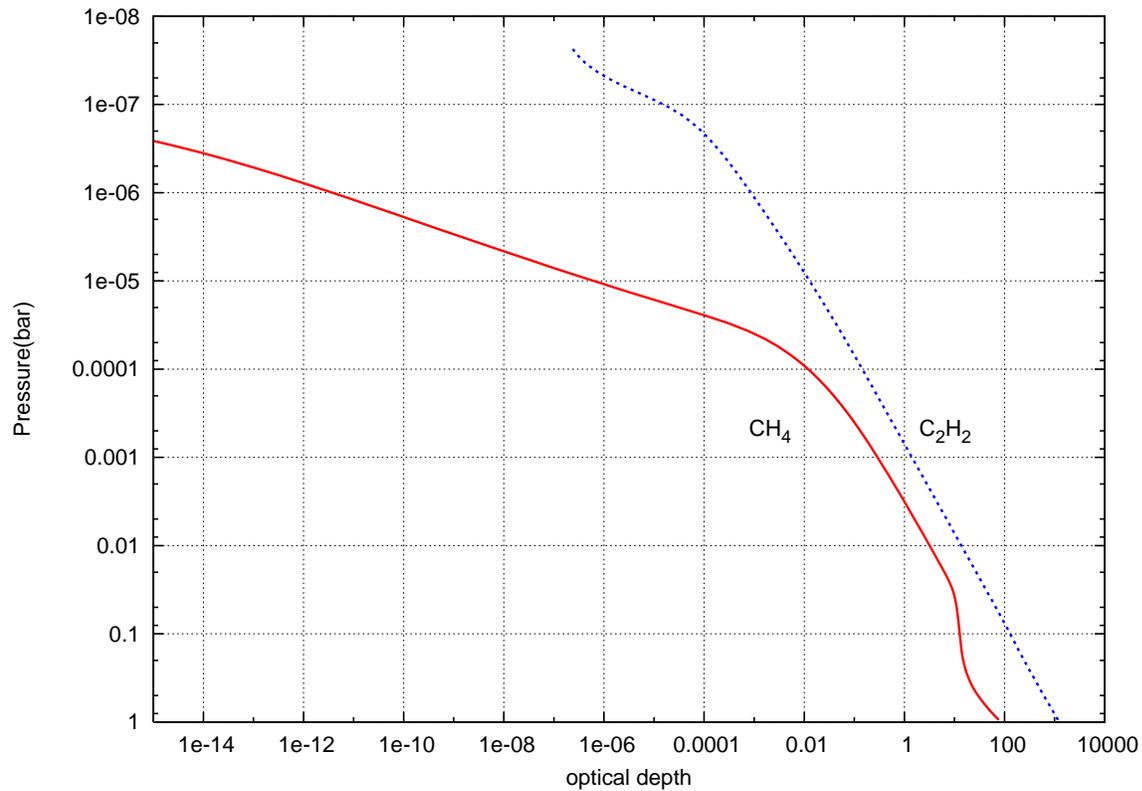}
\caption{Approximate optical depths of the $\ch4$ 7.7 micron band (red solid) and 
the 7.5 micron $\c2h2$ band (blue dashed) as a function of pressure.
The optical depth of $\c2h2$ is larger than $\ch4$ indicating that it may be the major absorber
in WASP-12b's atmosphere.}
\label{opticaldepth}
\end{figure}
\begin{figure}[!hbp|t]
\includegraphics[width=0.92\textwidth]{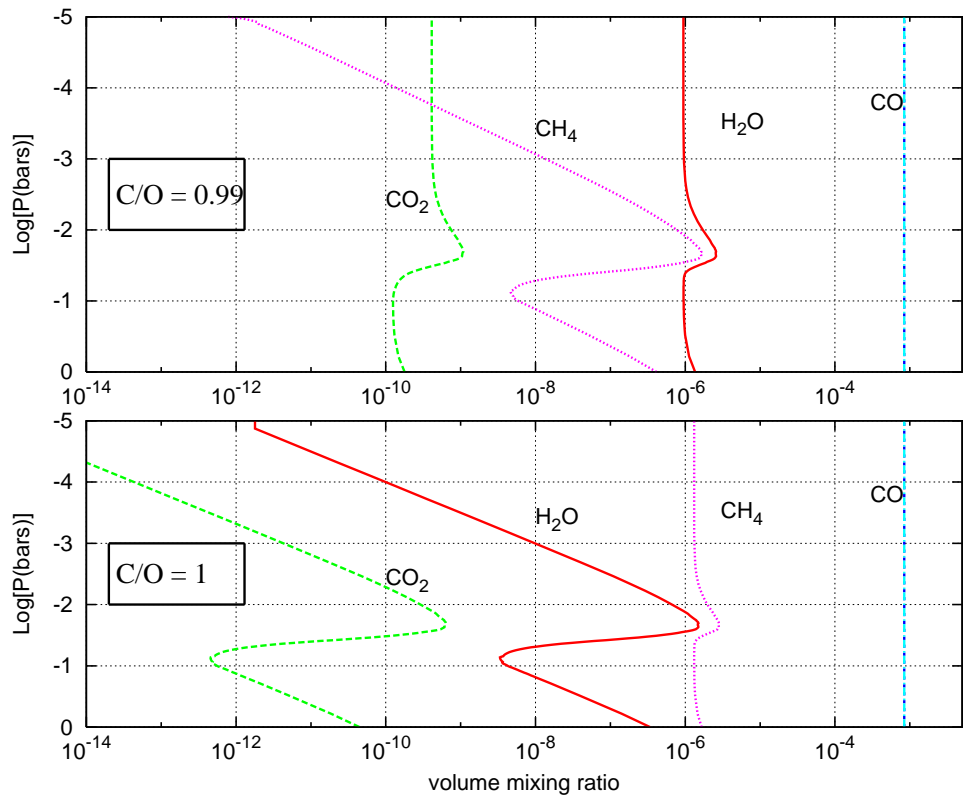}
\caption{
Diagrams showing the rapid shift in species' concentrations as C/O increases from 0.99
 (top panel) to 1.0 (bottom panel).}
\label{solarCO_transition}
\end{figure}
\begin{figure}[!hbp|t]
\includegraphics[width=0.92\textwidth]{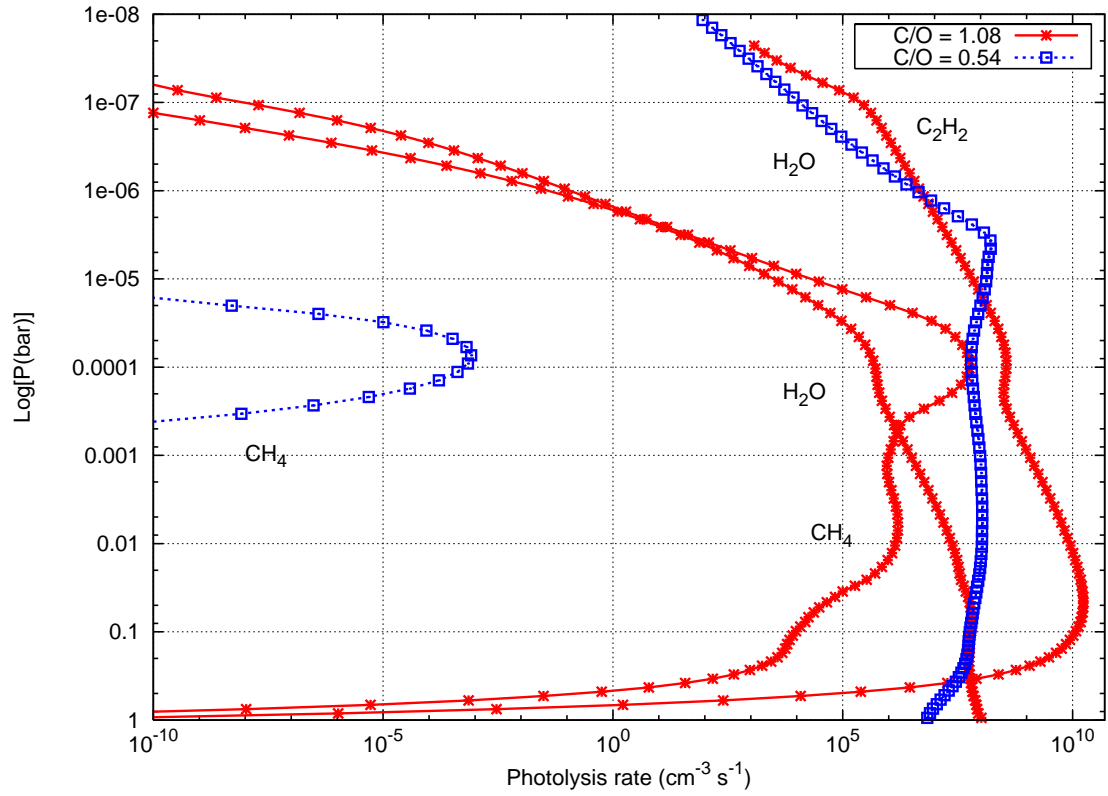}
\caption{Photolysis rates of $\h2o$ and $\ch4$ for $\ctoo = 0.54$ (dashed, solar) 
and $\ctoo = 1.08$ (solid, super-solar) including $\c2h2$. }
\label{solarphotorate}
\end{figure}

Based on their thermodynamic equilibrium calculations,  \cite{Madhu2011} concluded that Wasp-12b is abundant in $\ch4$ and deficient in $\h2o$.
Our analysis indicates that both the equilibrium and photochemical models predict $\c2h2$ and HCN are more abundant than $\ch4$. Also, 
$\c2h2$ has strong absorption in the range $2.98 - 3.1$ microns and also between $7.2 - 7.9$ 
microns, whereas
$\ch4$ has absorption features between $3.2 - 3.45$ microns and $7.3 - 8$ microns.
The short wavelength range for $\c2h2$  has little overlap with the Spitzer $3.6$ micron channel\footnote{Band pass
range from $3.08 - 4.01$ micron: \url {http://irsa.ipac.caltech.edu/data/SPITZER/docs/irac/calibrationfiles/spectralresponse/}}
 but the longer wavelength range for both  species overlaps with Spitzer's $8$ micron 
channel\footnote{Band pass range from $6.15 - 10.49$ microns:
 \url {http://irsa.ipac.caltech.edu/data/SPITZER/docs/irac/calibrationfiles/spectralresponse/}}. In order to determine which is the dominant
absorber, we have calculated the optical depths of the $\c2h2$ 7.5 micron band and the 
$\ch4$ 7.7 micron band as a function of pressure, 
as shown in Fig.\ref{opticaldepth}. Approximate band-averaged absorption coefficients for
 these features are $2 \times 10^{-19}$ cm$^{2}$ and $4 \times 10^{-19}$ cm$^{2}$,
respectively\footnote{\url {http://vpl.astro.washington.edu/spectra/c2h2pnnlimagesmicrons.htm}}.
The column depths are taken from our photochemical model. 
Clearly, $\c2h2$ has a larger optical depth than $\ch4$ and is
the dominant absorber. 
Note that, \cite{Madhu2011} point out that $0.01-1$ bar pressure levels contributes most to the
observed spectrum and that $\c2h2$ is considerably more abundant than $\ch4$ within this pressure
 range (Fig. \ref{solarCO1}).
Therefore, future analysis of
 observations of carbon rich planets (including further analysis of WASP-12b) should consider 
higher-hydrocarbon species.

\section{Discussion}
\label{discussion}
Our analysis confirms the previous thermodynamic equilibrium result that 
a $\ctoo \ge 1$ is needed to explain the observed overabundance of $\ch4$ in the atmosphere of
WASP-12b.
A similar conclusion
was reached by \cite{line2010} but for a different hot Jupiter planet, HD 189733b. These authors varied
$\ctoo$ from $0.1$ to $10$ times the solar value, while keeping the total metallicity at the solar 
value (Fig. 6 in their paper), and examined the effect on thermochemcal equilibrium mixing ratios at the 
lower boundary. As $\ctoo$ increases, most of the carbon in their model is in $\com$ and 
$\ch4$. At $\ctoo = 1$, $\h2o$ and $\ch4$ switch their profiles just as they do in our equilibrium 
models of WASP-12b (Fig. \ref{solarCO_transition}). Although our equilibrium models agree 
qualitatively with theirs, the respective mixing ratios of the major species differ because of different 
elemental abundances and overall hotter temperatures (their $\sim1500$ K versus our $2800$ K) 


We have also performed a sensitivity test to eddy diffusion 
varying  by 3 orders of magnitude above and below our eddy profile. For the larger case  the species
concentrations are well mixed over much of the atmosphere, deviating from the equilibrium even at 
relatively low altitudes  ($\sim 0.1$ bar). Consequently, the photolysis of $\c2h2$ and HCN is not effective
in producing atomic hydrogen (as mixing dominates photolysis even at high altitudes).
On the other hand, if the eddy diffusion coefficient is small (as proposed by \cite{YM2010}), 
mixing is not effective
and photochemistry becomes important at mid altitudes ($10^{-3} - 10^{-4}$ bar). Therefore,
 significant
deviations from equilibrium occur at all altitudes above this level.

\subsection{A possible mechanism for the origin of excess carbon in WASP-12b}

The  high $\ctoo$ ratio in WASP-12b is unexpected, considering that the host star has a solar
$\ctoo$ ratio (see \cite{Fossati2010} Table 2).  In the standard core accretion model
\citep{Pollack96},  volatiles such as carbon and oxygen are expected to remain unfractionated
in forming giant planets \citep{owen99}.
\cite{Lodders2004} pointed out that {\it Galileo} probe measurements of Jupiter's atmosphere show an
enriched carbon abundance of $1.7$ times solar and a depletion of oxygen by a factor of $4$ (but see
further discussion below).
To explain this result, \cite{Lodders2004} proposed a model in which 
carbonaceous matter began to condense in the solar nebula beyond $~5$ AU, thereby providing
the increased mass density needed for rapid core growth. By contrast, in the standard accretion model,
Jupiter forms just beyond the ``ice line'' where water ice begins to condense. In the \cite{Lodders2004}
model, the ice line would have been farther out, beyond the orbit of Jupiter, and this would
explain Jupiter's apparent deficiency in O relative to C. A similar mechanism might then account for
the high $\ctoo$ ratio in  WASP 12b.

Although the \cite{Lodders2004} model could be correct, we suspect that Jupiter formed beyond the ice
line, with a solar $\ctoo$ ratio, and that other factors are responsible 
for observed $\ctoo$ enrichments in exoplanets. The Galileo Probe is widely thought to have descended into
an infrared ``hot spot'' \citep{atreya99}, that is, 
an area of downwelling air that had been depleted in $\h2o$ during its uplift from below. In support of this
idea, the  $\h2o$ mixing ratio was observed to gradually increase with depth down to $20$ bars, at which
point the probe lost contact with Earth (see \cite{atreya99} Table 1).
 Furthermore, thunderstorms and lightining were also observed by the probe deeper than
$4-5$ bars \citep{Gierasch2000, Ingersoll2000, atreya2005}, which is consistent with equilibrium
cloud condensatation models which predict that water clouds can form in this pressure range if the oxygen
abundance is at least solar \citep{atreya2005} \footnote{It should be
noted that, because the base level of the water clouds was not determined, the water abundance in
the deep well-mixed regions of Jupiter is still unknown. However, this does not change the observed result
that the mixing ratio of $\h2o$ gradually increases with depth.}.

As an alternative to the \cite{Lodders2004} model, we suggest that the high $\ctoo$ for WASP-12b  
arose because the primordial disk was depleted in oxygen abundance during the giant planet's migration.
The carbon compounds ($\ch4, \com$) may have been trapped in ices in the form of planetesimals and then
accreted onto the envelope of the gas giant, resulting in the observed enhancement of $\ctoo > 1$. 
Assuming that the disk started with solar elemental abundances  of carbon
 ($2.26 \times 10^{-4}$) and oxygen ($4.20 \times 10^{-4}$), in order to obtain $\ctoo = 1.08$ 
in WASP-12b (our high $\ctoo$ model case), the
 [O] abundance in the disk must have been depleted by $\approx 50 \%$.
Recently, \cite{Madhu2011b} performed a more detailed analysis of the formation of WASP-12b and 
concluded that the primordial disk was depleted in [O]  by $41 \%$. The discrepancy
in our numbers arises because \cite{Madhu2011b} used elemental abundances of the host star WASP-12
\citep{Fossati2010}, which are $3.54 \times 10^{-4}$ and $7.94 \times 10^{-4}$ for [C] \& [O],
respectively.

It is possible that the depletion of [O] in WASP-12b occured because the host star accreted fractionated refractory
 materials (that trapped $41 \%$ of [O], in the WASP-12 case)
 from the protoplanetary disk during planetary formation.
In our Solar System, planetary migration 
could have affected giant planet composition to some extent, as Jupiter and Saturn, in particular, are
thought to have moved around considerably during planetary accretion
\citep{Tsiganis2005, Gomes2005, Morbidelli2005, Walsh2011},
with Jupiter perhaps coming as close in as 1.5 AU.
 But, as far as we know, giant planets never migrated through the terrestrial planet region of our
system. By contrast, in the WASP-12 system,
planetary migration was evidently much more pronounced. WASP-12b migrated from the outer parts of the
nebula to its present location close to the star. Meanwhile, rocky planets formed in the hot, inner
parts of the nebula may have migrated in even closer and been consumed by the star. If WASP-12b
accreted additional material during its journey, that material would have been depleted in O relative
to C, possibly accounting for the high $\ctoo$ ratio of the planet. \cite{Madhu2011b} mention this 
possibility, but they rule it out because they argue  this would require that [C]/[H] in the envelope
of WASP-12b should be close to the host star's value, which is not the case. We don't agree with this
objection, however, because all known planets accrete elements heavier than He more efficiently than 
they do H. Jupiter, for example, is enriched in C, N, and S compared to the Sun by a factor of ~3 
\citep{beatty1999}.
Further observations may be needed to determine the validity
of this mechanism.

The accretion of refractory elements onto a star has been proposed as one of the reasons why
solar twins and analogs in the solar neighborhood have enhanced heavy elemental abundances compared to the Sun
\citep{Melendez2009, Ramirez2009}. These studies found that the abundances of heavy elements in these solar
analog stars increase with their condensation temperature. The authors attribute the
apparent depletion of refractory elements in the Sun to the existence of terrestrial planets,  and
they suggest
that Sun accreted refractory-depleted material from the nebula during the formation of the 
solar system\footnote{Note
that the accreted material is not depleted in elements such as carbon because they have
low condensation temperatures and so did not condense in the inner part of the solar nebula.}.
 Indeed,
\cite{Chambers2010} showed that  adding $4 \mearth$ of Earth-like and
carbonaceous-chondrite-like material  to the solar convection zone brings the Sun's elemental abundance
 in line with the mean abundances of solar twins.

Recently, \cite{schuler2011} derived precise elemental abundances for ten stars using high-resolution
 spectroscopy.  All ten of their stars have at least one giant planet around them at different orbital
distances. Their analysis
indicated that four stars, all of which have hot Jupiters ($\sim 0.05$ AU), show positive correlations between
refractory elemental abundance and condensation temperature. This implies
that these stars may have accreted refractory-rich planetary material or cores.
 If a similar accretion happened during the formation of the WASP-12 system, then ``pollution'' 
signatures
in the atmosphere of WASP-12 may be observable.
 Formation models of  protostars
from molecular clouds \citep{WT2003} indicate that solar-mass protostars have thin convective 
envelopes
($\sim 0.02 $ M$_{\odot}$),
similar to the present day Sun, and hence mixing of deposited material may not be significant.
But it has been suggested that WASP-12b may be losing mass to its star \citep{Li2010}. If
this is the case, then it could be difficult to determine  how much of the refractory
material has accreted onto  WASP-12 during its formation.
  \cite{Fossati2010} performed a detailed
spectropolarimetric analysis of WASP-12 to look for pollution signatures due to the material lost
by WASP-12b. They found  hints of pollution but were unable to draw firm conclusions.
 A differential
analysis of WASP-12 twins (with the same effective temperature, age and metallicity), identifying
their abundances with high precision, is required to determine if the
refractory elemental abundance of WASP-12 does indeed increases with condensation temperature.

\section{Conclusions}
\label{conclusions}

In this study, we analyzed  how a disequilibrium mechanism such as photochemistry
can affect the observed abundances  of  $\h2o$, $\com$,
$\co2$ and $\ch4$ in the WASP-12b atmosphere. We considered two models, with $\ctoo = 0.54$
(solar)
 and $\ctoo = 1.08$ (twice solar). 
Although our photochemical results agree that 
high $\ctoo$ is needed to explain the observed high abundance of $\ch4$ and
 lack of observable $\h2o$, they also indicate $\c2h2$ and HCN are more abundant than $\ch4$ and
should be taken into consideration in modeling hot-jupiter atmospheres\footnote{Note that this prediction is not specifically a result of
our photochemical model, as our thermodynamic equilibrium models predict this, as well.}. More
importantly, our results indicate that $\c2h2$ is the dominant absorber  at 1.6 and 8 micron
 in WASP-12b's  atmosphere and the absorption features may possibly be arising from $\c2h2$ rather than
$\ch4$. Observations with Hubble Space Telescope's WFC3 can resolve this discrepancy.


We also propose a possible mechanism for the origin of the excess carbon observed in WASP-12b.  
Following other authors, we suggest that WASP-12 may have
accreted rocky, O-rich material from the nebula during the formation of the system, leaving
the disk relatively enriched in other volatiles such as C and N. WASP-12b then accreted some of this
high $\ctoo$ material, which thereby gave rise to the high $\ctoo$ ratio of the planet. Testing this 
hypothesis
requires that we understand whether WASP-12 is currently stealing mass from its planet, WASP-12b. A high
precision abundance analysis of WASP-12 twins and analogs can shed light 
on the refractory elemental abundance of WASP-12 and the 
possible origin of excess carbon in WASP-12b.

\acknowledgements

The authors would like to thank the referee, Julianne Moses, for pointing to us the importance of
$\c2h2$ and HCN chemistry and for her in-depth  analysis of our work 
which helped in improved  photochemical models and our current manuscript.
 R. K and J.F.K gratefully acknowledge funding from NASA Astrobiology Institute's  Virtual 
Planetary Laboratory lead team, supported by NASA under cooperative agreement
NNH05ZDA001C, and the Penn State Astrobiology Research Center.

\clearpage
\thispagestyle{empty}

\clearpage
\thispagestyle{empty}

\clearpage
\thispagestyle{empty}

\clearpage
\thispagestyle{empty}

\clearpage
\thispagestyle{empty}

\clearpage
\thispagestyle{empty}

\clearpage
\thispagestyle{empty}


\clearpage
\appendix
\section{Supplementary figures}
\label{supp}


\thispagestyle{empty}
\begin{figure}[!hbp|t]
\includegraphics[width=0.92\textwidth, angle=0]{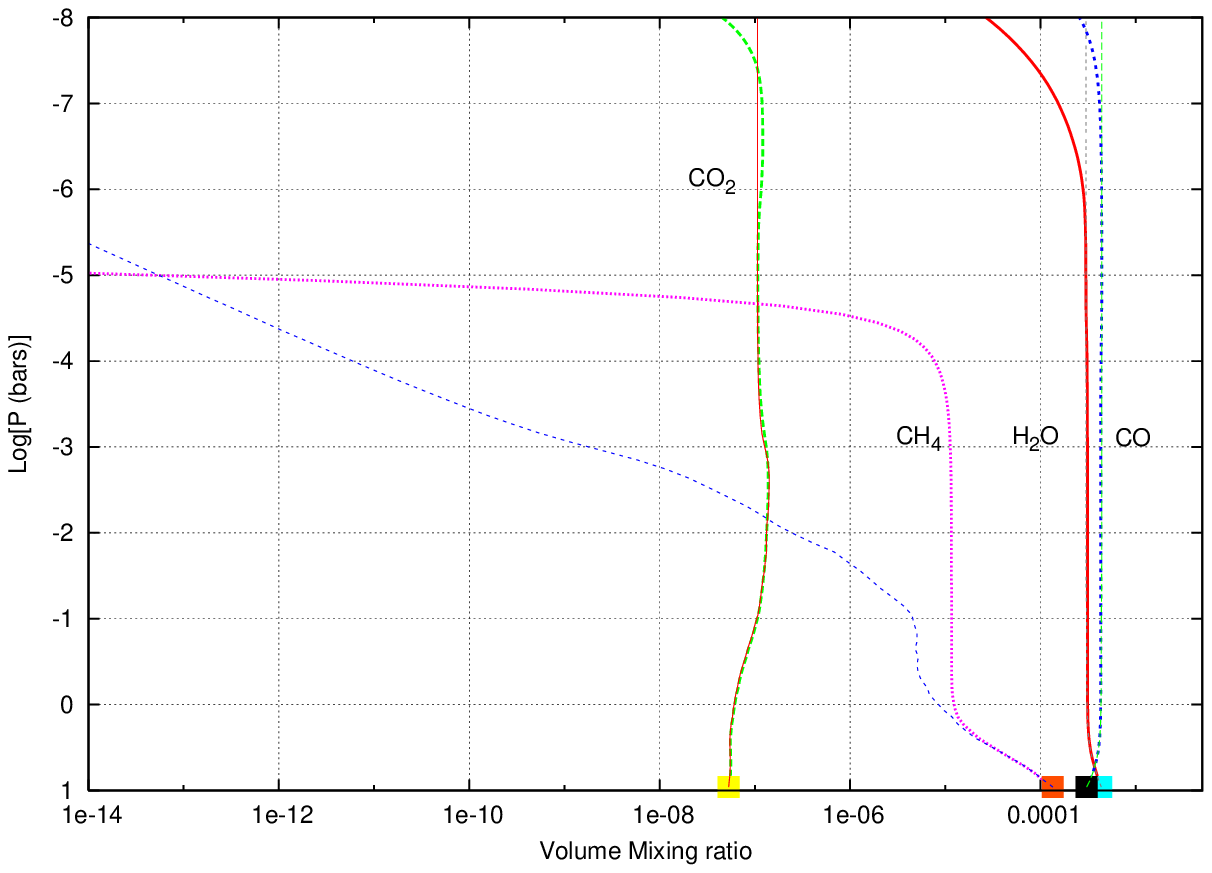}
\caption{Photochemical  mixing ratio profiles of $\h2o, \com, \ch4$ and 
$\co2$ derived from our model of the dayside atmosphere of HD 189733b. For comparison,
 see \cite{line2010} and \cite{Moses2011}.  }
\label{hd189733}
\end{figure}

\begin{deluxetable}{ccccccc}
\tablecaption{Reaction list and rate constants \citep{Zahnle2011} used in this study. Only the forward rate constants are given as
we calculate reverse rate constants from the forward rate assuming thermodynamic equilibrium (see discussion in Section 2). For
three body reactions, the first and second row represent the low and high pressure rate limits, respectively.
 A full version of
 the table is available in the electronic edition of the {\it Astrophysical Journal}.}
\tablewidth{0pt}
\tablehead{ \colhead{Number} & \colhead{Reactants} &\colhead{} &\colhead{Products}  & \colhead{Rate\tablenotemark{a}\tablenotetext{a}{2-body reaction rates are in cm$^{3}$~s$^{-1}$; 3-body rates are in cm$^{6}$~s$^{-1}$.}} & \colhead{ Reference}}   
\startdata  $1$ & H $+$ H $+$ M & $\rightarrow$ &  H$_{2}+$  M & $8.8 \times 10^{-33} (T/298)^{-0.60}$ & Baulch et al.(1992) \\ 
& H $+$ H   & $\rightarrow$ & H$_{2}$  & $1.0 \times 10^{-12}$ &\\
$2$ & O $+$ H $+$ M & $\rightarrow$ & OH $+$ M & $4.3 \times 10^{-32}$ & Tsang \& Hampson (1986)\\
& O $+$ H & $\rightarrow$ & OH  & $1.0 \times 10^{-12}$ &\\
$3$ & H$_{2}$ $+$ O & $\rightarrow$ & OH $+$ H & $3.5 \times 10^{-13} (T/298)^{2.67} e^{-3160/T}$ & Baulch et al.(1992)  \\ 
$4$ & H$+$ OH $+$ M & $\rightarrow $ & H$_{2}$O $+$  M & $6.6 \times 10^{-32} (T/298)^{-2.1} $ & Javoy et al.(2003)  \\ 
& H$+$ OH & $\rightarrow $ & H$_{2}$O & $2.7 \times 10^{-10} e^{-75/T}$ & Cobos \& Troe (1985)\\
$5$ & H$_{2} +$ OH & $\rightarrow$ & H$_{2}$O $+$ H & $1.6 \times 10^{-12} (T/298)^{1.60} e^{-1660/T}$ & Baulch et al.(1992)  \\ 
\\
\enddata
\label{table1}
\end{deluxetable}

\end{document}